# Detecting a Light Gravitino at Linear Collider to Probe the SUSY Breaking Scale

Daniel R. Stump, Michael Wiest and C.–P. Yuan

*Department of Physics and Astronomy, Michigan State University,*

*East Lansing, MI 48824, USA*

**Abstract**

If supersymmetry is dynamically broken at a low scale ($M_{susy}$), within a few orders of magnitude of the weak scale, then the lightest supersymmetric partner is the gravitino and the next to lightest supersymmetric partner is a neutralino $\chi_1^0$ with mass $m_{\chi_1^0}$, which can decay into a photon ($\gamma$) plus a gravitino ($\widetilde{G}$). We study the detection of $e^-e^+ \to \chi_1^0 \chi_1^0 \to \gamma \widetilde{G} \gamma \widetilde{G}$ at the proposed Linear Collider, and find the range of the parameters $M_{susy}$ and $m_{\chi_1^0}$ that can be accessible with a right-hand polarized electron beam at $\sqrt{S} = 500\,\text{GeV}$, with $50\,\text{fb}^{-1}$ integrated luminosity. We also discuss briefly the accessible range for current electron and hadron colliders.

# 1  Introduction

Recently, Dine, Nelson and Shirman [1] have presented a model in which supersymmetry (SUSY) is dynamically broken at a low scale, within a few orders of magnitude of the weak scale. They noted that in such a model the lightest supersymmetric partner (LSP) is the gravitino ($\widetilde{G}$) and the next to lightest supersymmetric partner (NLSP) is a neutralino $\chi_1^0$, which is a mixture of neutral gauginos and Higgsinos. One decay mode of the NLSP, the decay mode considered in this paper, is the decay into a photon and a light gravitino. The mass of the gravitino $m_{3/2}$ is in the range $1\,\text{eV} < m_{3/2} < 10\,\text{KeV}$. The supersymmetry-breaking scale is $M_{susy} = \sqrt{m_{3/2} M_{Planck}}$, where the Planck mass is $M_{Planck} = 1/\sqrt{G_N} \sim 10^{19}\,\text{GeV}$; hence, $M_{susy}$ is in the range $10^5 - 10^7\,\text{GeV}$. This model differs from most other models of supersymmetry breaking, such as the minimal supersymmetric standard model (MSSM), in which the gravitino mass $m_{3/2}$ is of order the weak scale (*i.e.*, a few hundred GeV to 1 TeV) and the supersymmetry-breaking occurs at a scale $M_{susy} = \sqrt{m_{3/2} M_{Planck}}$ in the range $10^{10} - 10^{11}\,\text{GeV}$. Usually, in the MSSM, the lightest supersymmetric partner is a neutralino, which is then stable assuming *R*-parity conservation. Thus the decay of the neutralino into a photon and a light gravitino is a distinctive model-independent signal for low-energy supersymmetry breaking.

Motivated by Ref. [1], we examine in this paper the detection of a light gravitino from the decay of massive neutralinos (the NLSP, with mass $m_{\chi_1^0}$) which are produced in pairs at Linear Collider (a proposed future $e^- e^+$ collider) with a right-hand polarized electron beam. (Using a right-hand polarized beam is to suppress background contributions, as discussed in Sec. 2.4.) For this study, we assume that the center-of-mass energy $\sqrt{S}$ of the Linear Collider (LC) is $500\,\text{GeV}$ and its luminosity is $50\,\text{fb}^{-1}$ per year. Accordingly, we ignore the mass of the electron and the mass of the gravitino in this study. The experimental signature of the signal event of interest is two photons with large missing energy. The missing energy belongs to the two gravitinos, which escape undetected. We shall discuss the range of the parameters $M_{susy}$ and $m_{\chi_1^0}$ that can be accessible at LC. The accessible parameter range for current electron (LEP/SLC) and hadron (Tevatron) colliders is also discussed.

## 2 Detecting a Light Gravitino at Linear Collider

### 2.1 Production of NLSP Pairs at LC

A pair of NLSP's can be produced at LC via the tree-level process $e^-e^+ \to \chi_1^0 \chi_1^0$. At tree level, the particles exchanged in the t- and u-channel Feynman diagrams are either left or right selectrons (the scalar superpartners of left-handed or right-handed electrons, respectively), and that of the s-channel diagram is the $Z$-boson. The t- and u-channel diagrams vanish if $\chi_1^0$ is Higgsino-like because the coupling of Higgsino-electron-selectron is zero for a massless electron. The s-channel diagram vanishes if $\chi_1^0$ is gaugino-like because there is no tree-level $Z$-$\widetilde{B}$-$\widetilde{B}$ or $Z$-$\widetilde{W^3}$-$\widetilde{W^3}$ interaction, by the $SU(2)_L \times U(1)_Y$ symmetry. ($\widetilde{B}$ is the supersymmetric partner of the $U(1)_Y$ gauge boson $B$, and $\widetilde{W^3}$ is the supersymmetric partner of the third component of the $SU(2)_L$ gauge boson $W^3$.)

In Ref. [2], it was shown that in the constrained MSSM (CMSSM), the lightest neutralino $\chi_1^0$ is most likely to be a $\widetilde{B}$, *i.e.,* is $\widetilde{B}$-like for the largest allowed region of model space, although in general it is a mixture of the neutral gauginos ($\widetilde{B}$ and $\widetilde{W^3}$) and neutral Higgsinos. If $\chi_1^0$ is found, it is important to determine its components of gauginos and Higgsinos in order to test different models of supersymmetry-breaking. Fortunately, LC will provide a powerful tool to achieve this goal. The electron beam at LC can be highly polarized – more than 90% polarization is expected [3]. For simplicity, in this paper, we shall assume that the electron beam at LC can be 100% right-hand polarized.

(The assumption of 100% polarization will not significantly affect our conclusions as compared to a 90% polarized beam.) In this case, the $\widetilde{W^3}$ component of $\chi_1^0$ does not contribute to $\chi_1^0\chi_1^0$ production in $e_R^-e^+$ collisions because $\widetilde{W^3}$ couples only to a left-handed electron, due to the $SU(2)_L$ gauge symmetry. As for the $\widetilde{B}$ component, recall that $\widetilde{B}$ may be written as a linear combination of photino $\widetilde{\gamma}$ and $\widetilde{W^3}$; specifically,

$$\widetilde{B} = \widetilde{\gamma}/\cos\theta_w - \widetilde{W^3}\tan\theta_w , \tag{1}$$

where $\theta_w$ is the weak mixing angle ($\sin^2\theta_w = 0.23$). Therefore, if $\chi_1^0$ is gaugino-like, then the production of $\chi_1^0\chi_1^0$ from a right-hand polarized electron beam occurs only through the $\widetilde{\gamma}$ component in Eq. (1). In Secs. 2.1 to 2.4 we assume that the neutralino $\chi_1^0$ is gaugino-like; in Sec. 2.5 we discuss the case that $\chi_1^0$ is Higgsino-like. For a gaugino-like neutralino $\chi_1^0$ we define a mixing angle $\alpha$ by

$$\chi_1^0 = \widetilde{B}\cos\alpha + \widetilde{W^3}\sin\alpha . \tag{2}$$

Then the cross-section for $\chi_1^0\chi_1^0$ production from a right-handed electron is, by Eqs. (2) and (1),

$$\sigma(e_R^-e^+ \to \chi_1^0\chi_1^0) = \sigma(e_R^-e^+ \to \widetilde{\gamma}\widetilde{\gamma}) \left(\frac{\cos\alpha}{\cos\theta_w}\right)^4 \tag{3}$$

in terms of the cross-section for photino-pair production. In the numerical calculations below we always report results for $\alpha = 0$, corresponding to $\chi_1^0$ being $\widetilde{B}$-like; it is simple to determine the numbers for other values of $\alpha$.

The cross-section for $e_R^-e^+ \to \widetilde{\gamma}\widetilde{\gamma}$ has been published [4, 5]. There are two Feynman diagrams for this process, with t-channel and u-channel exchange of a right selectron $\widetilde{e_R}$. The differential cross-section $d\sigma/d\cos\theta$ for $e_R^-(p_1) + e^+(p_2) \to \chi_1^0(p_3) + \chi_1^0(p_4)$, where $\theta$ is the angle of an outgoing neutralino $\chi_1^0$ relative to the $e^-$ direction, is

$$\frac{d\sigma}{d\cos\theta} = \frac{e^4}{32\pi S}\sqrt{1-\frac{4m_{\chi_1^0}^2}{S}}\left\{\left(\frac{t-m_{\chi_1^0}^2}{m_{\tilde{e}_R}^2-t}\right)^2 + \left(\frac{u-m_{\chi_1^0}^2}{m_{\tilde{e}_R}^2-u}\right)^2\right.$$

$$\left.-\frac{2Sm_{\chi_1^0}^2}{(m_{\tilde{e}_R}^2-t)(m_{\tilde{e}_R}^2-u)}\right\}\left(\frac{\cos\alpha}{\cos\theta_w}\right)^4 \quad (4)$$

where $t = (p_1 - p_3)^2$ and $u = (p_1 - p_4)^2$. This cross-section is shown in Fig. 1 for $\sqrt{S} = 500\,\text{GeV}$, for $\alpha = 0$ and two cases of the mass parameters: $m_{\chi_1^0} = 100\,\text{GeV}$, $m_{\tilde{e}_R} = 300\,\text{GeV}$, and $m_{\chi_1^0} = 200\,\text{GeV}$, $m_{\tilde{e}_R} = 600\,\text{GeV}$.

## 2.2 Decay of the Neutralino

Since the photino component of the neutralino $\chi_1^0$ would mainly decay into a photon and a gravitino for models [1] in which the gravitino is extremely light, the experimental signature of the signal event we consider in this paper is two photons and missing energy. The missing energy belongs to the two gravitinos. Other signals could be considered that involve the zino $\tilde{Z}$ component of $\chi_1^0$, which would decay to a $Z$-boson and a gravitino; these signatures are $\gamma Z$ with missing energy and $ZZ$ with missing energy.

The rate of the neutralino decay $\chi_1^0 \to \gamma\tilde{G}$ can be related to the rate of the photino decay $\tilde{\gamma} \to \gamma\tilde{G}$ from the fact that $\chi_1^0$ is a mixture of $\tilde{\gamma}$ and $\tilde{Z}$

$$\chi_1^0 = \tilde{\gamma}\cos(\alpha - \theta_w) + \tilde{Z}\sin(\alpha - \theta_w)\,, \quad (5)$$

and the fact that at tree level only the $\tilde{\gamma}$ component decays to $\gamma\tilde{G}$, with nearly 100% branching ratio. Therefore we have

$$\Gamma(\chi_1^0 \to \gamma\tilde{G}) = \Gamma(\tilde{\gamma} \to \gamma\tilde{G})\cos^2(\alpha - \theta_w)\,. \quad (6)$$

The photino decay rate has been calculated [6, 4] to be

$$\Gamma(\tilde{\gamma} \to \gamma\tilde{G}) = \frac{m^5}{8\pi d^2}\,, \quad (7)$$

where $m$ is the photino mass, assuming the photino is a mass eigenstate. In general, however, $\widetilde{\gamma}$ is not a mass eigenstate, and in Eq. (6) we use the neutralino mass $m_{\chi_1^0}$; that is,

$$\Gamma(\chi_1^0 \to \gamma \widetilde{G}) = \frac{m_{\chi_1^0}^5}{8\pi d^2} \cos^2(\alpha - \theta_w) \ . \tag{8}$$

The parameter $d$ is related to the scale of supersymmetry breaking [7] by

$$d = \frac{\sqrt{3}}{\sqrt{4\pi}} m_{3/2} M_{Planck} = \frac{\sqrt{3}}{\sqrt{4\pi}} M_{susy}^2 \ . \tag{9}$$

The decay rate in Eq. (7) is derived by using the Goldstino component of the gravitino in the $\widetilde{\gamma}$-$\gamma$-$\widetilde{G}$ vertex factor, which is then $[\not{p}_\gamma, \gamma^\mu] \not{p}_{\widetilde{G}}/2d$ [7]. Note that the decay rate is inversely proportional to the fourth power of the supersymmetry-breaking scale $M_{susy}$ and is proportional to the fifth power of the neutralino mass. For $M_{susy} = 10^6$ GeV and $m_{\chi_1^0} = 100$ GeV, which are characteristic values for our study, the lifetime of the NLSP $\chi_1^0$, given by

$$\tau = \frac{1}{\Gamma} = \frac{6 M_{susy}^4}{m_{\chi_1^0}^5} \frac{1}{\cos^2(\alpha - \theta_w)} \ , \tag{10}$$

is expected to be $5 \times 10^{-10}$ sec for $\alpha = 0$. For $M_{susy}$ in the range $10^5 - 10^7$ GeV, $\tau$ can vary from $10^{-14}$ sec to $10^{-6}$ sec.

The typical distance travelled by a neutralino before it decays at LC is

$$D = \gamma \beta c \tau \ , \tag{11}$$

where $\beta$ is the neutralino velocity $\beta = \sqrt{1 - 4m_{\chi_1^0}^2/S}$ and $\gamma = 1/\sqrt{1-\beta^2}$. One must observe the decay photon to make any statement about existence of a light gravitino, and to detect the photon from $\chi_1^0 \to \gamma \widetilde{G}$ the decay must occur within the detector volume. We shall assume that $D$ must be smaller than 1 meter to observe the decay photon inside the detector. Figure 2, which is explained further below, shows the range of parameters $M_{susy}$ and $m_{\chi_1^0}$ that can be accessible at LC.

## 2.3 Background-Free Signal Events

Because of the distinct signature of the signal event, two photons plus missing energy, there will be no background to the signal if the decay photons (with high transverse momenta) can be pointed back to their parents (massive $\chi_1^0$'s) and it can be verified that they do not come from the interaction point (IP) of the collider. At LC the interaction region may be considered to be a point, because the beam sizes are very small ($\sigma_x \times \sigma_y \times \sigma_z = 5\,\mathrm{nm} \times 300\,\mathrm{nm} \times 100\,\mu\mathrm{m}$) and this IP is designed to remain stable at about the same scale [8]. Depending upon calorimeter technology, the angular resolution for the trajectory of a few GeV photon is typically in the range 10 to 100 mrad, assuming the photon originates within the inner portion of the calorimeter. Given a 1 m radius for the calorimeter, the pointing resolution to the IP would be 1 cm to 10 cm. Note that the resolution will generally improve as $1/\sqrt{E_\gamma}$. The typical energy $E_\gamma$ of the photon from the decay of a massive $\chi_1^0$ is $\sqrt{S}/4$; more precisely, the distribution in photon energy is constant with $E_\gamma$ between $\sqrt{S}(1-\beta)/4$ and $\sqrt{S}(1+\beta)/4$, as shown in Fig. 3. To be conservative, we will assume in this paper that if $\chi_1^0$ travels more than 10 cm before it decays (*i.e.*, $D \geq 10\,\mathrm{cm}$), then the displaced origin of the decay photon can be well-seperated from the IP at LC.

If $10\,\mathrm{cm} \leq D \leq 1\,\mathrm{m}$, then this signal event is background-free. The range of $M_{susy}$ and $m_{\chi_1^0}$ in which this condition on $D$ is satisfied is the cross-shaded region in Fig. 2. We see that for $M_{susy} = 10^6\,\mathrm{GeV}$, the background-free range of $m_{\chi_1^0}$ is from 85 GeV up to 122 GeV. (We note that for fixed $M_{susy}$ and $\sqrt{S}$, $D$ is proportional to $\beta/m_{\chi_1^0}^6$.) For $m_{\chi_1^0} = 100\,\mathrm{GeV}$, the background-free signal occurs for $M_{susy}$ around $10^6\,\mathrm{GeV}$. Since we are considering a signal with no background, observing just one event would be sufficient to declare the discovery of the signal. The integrated luminosity of LC is $50\,\mathrm{fb}^{-1}$ per year, so an LC experiment is sensitive to a signal production cross-section larger than 0.02 fb. In our case, because the electron beam is right-hand polarized, the production of $\chi_1^0\chi_1^0$ occurs by exchange of a right selectron $\widetilde{e}_R$, and the cross-section depends on the mass of $\widetilde{e}_R$. (In this work, we shall ignore the possible small mixing between $\widetilde{e}_R$ and $\widetilde{e}_L$ so that $\widetilde{e}_R$ is the mass eigenstate with mass $m_{\widetilde{e}_R}$.) In the model [1] we are considering, the LSP is the gravitino and the NLSP is $\chi_1^0$, so $\widetilde{e}_R$ must be heavier than $\chi_1^0$. Also, in this model, we expect $m_{\widetilde{e}_R}$ to be of the same order as $m_{\chi_1^0}$, because sfermion masses squared ($m_{\widetilde{e}_R}^2$) depend on a two-loop diagram while gaugino masses ($m_{\chi_1^0}$) depend on a one-loop diagram. (The details depend on the gauge groups assumed in the model.) In Table 1 we give the signal production cross-section $\sigma$ for a few choices of $m_{\chi_1^0}$ and $m_{\widetilde{e}_R}$ at LC, assuming $M_{susy} = 10^6$ GeV. In all cases the cross-section is larger than 1 fb. The typical decay length $D$ is also given in Table 1. We conclude that for $M_{susy} = 10^6$ GeV and $m_{\widetilde{e}_R}$ less than 1 TeV, LC guarantees the discovery of the background-free signal event if $85 \, \text{GeV} \leq m_{\chi_1^0} \leq 122 \, \text{GeV}$. In fact, it is the decay length $D$, not the production cross-section $\sigma$, that constrains the discovery of the background-free signal event, because $\sigma$ is larger than 0.02 fb for $m_{\chi_1^0} < \sqrt{S}/2 = 250 \, \text{GeV}$ and $m_{\widetilde{e}_R} < 1 \, \text{TeV}$ except when $m_{\chi_1^0}$ is extremely close to $\sqrt{S}/2$. Our conclusion does not change even if we require observation of at least 10 signal events to declare discovery of the signal.

In Fig. 2, the shaded region shows the values of $M_{susy}$ and $m_{\chi_1^0}$ for which $D \leq 1 \, \text{m}$ and for which the number of events at LC ($\sigma \times 50 \, \text{fb}^{-1}$) is greater than 10. The diagonal boundary on the left is the curve $D = 1 \, \text{m}$, and the vertical boundary on the right is where the number of events equals 10. The cross-section $\sigma$ does not depend on $M_{susy}$ so the boundary on the right side is a vertical line near the threshold at $m_{\chi_1^0} = \sqrt{S}/2 = 250$ GeV. Figure 2 is for $m_{\widetilde{e}_R} = 300 \, \text{GeV}$, where the right selectron $\widetilde{e}_R$ is the exchanged particle in the process $e_R^- e^+ \to \chi_1^0\chi_1^0$. Increasing $m_{\widetilde{e}_R}$ up to 1 TeV only slightly shifts the vertical boundary on the right toward the left. The cross-shaded region has 10 cm $\le D \le 1$ m, corresponding to the background-free signal. The two diagonal boundary curves depend only on the kinematics, not on the cross-section, so they do not depend on the selectron mass $m_{\widetilde{e}_R}$. In the rest of the shaded region, *i.e.*, not including the cross-shaded region, we have $D \le 10$ cm, for which we assume conservatively it is not possible to detect a displaced vertex of the decay photon. For these values of $M_{susy}$ and $m_{\chi_1^0}$ there are intrinsic backgrounds to the signal process, which we discuss next.

## 2.4 Non-Background-Free Signal Events

As shown in Fig. 2 and Table 1, assuming $M_{susy} = 10^6$ GeV, if $m_{\chi_1^0} \ge 120$ GeV then the decay length $D$ is less than 10 cm, which does not satisfy the criterion for being a background-free signal event. The experimental signature of the signal event in this case consists of two photons coming out of the interaction region with large missing energy. We must consider background processes, *i.e.*, standard processes with a similar event signature. First, let us examine the characteristic features of the signal event. To obtain the distributions of the decay photons we evaluated the full correlated helicity amplitudes including the decays of the two neutralinos. However, we did not find a noticeable difference compared to a simpler calculation in which the decays of the $\chi_1^0$'s are treated independently, ignoring the correlation between the polarizations of the two $\chi_1^0$'s. Figure 3 shows the distribution of the single-photon energy $E_\gamma$, for $m_{\chi_1^0} = 100$ GeV and $m_{\widetilde{e}_R} = 300$ GeV. The distribution is approximately constant with $E_\gamma$ between $\sqrt{S}(1-\beta)/4$ and $\sqrt{S}(1+\beta)/4$, so the order of magnitude of the photon energies is $\sqrt{S}/4$. Typically, each photon has large transverse momentum $p_T^\gamma$. Figure 4 shows the $p_T^\gamma$ distribution of either photon, for $m_{\chi_1^0} = 100$ GeV and $m_{\widetilde{e}_R} = 300$ GeV. If $\chi_1^0$ is heavy, then the two decay photons are acollinear, which indicates missing transverse momentum. If $\chi_1^0$ is light, then due to the large momenta of the $\chi_1^0$'s the two photons will tend to be more nearly back-to-back, but the sum of the two photon energies will still be peaked at about $\sqrt{S}/2$, which indicates missing energy. This latter feature of the signal event is shown in Fig. 5, which is the distribution of the sum of photon energies, for $m_{\chi_1^0} = 100\,\text{GeV}$ and $m_{\widetilde{e}_R} = 300\,\text{GeV}$. (Figures 3–5 are for $\alpha = 0$.)

Since the signal event consists of two photons, the first obvious background process to consider is the ordinary QED process $e_R^- e^+ \to \gamma\gamma$. However, the photon kinematics for this process are much different than for the signal process: The photons will be approximately back-to-back with combined energy $\sqrt{S}$. Initial state radiation can change the energy by a small amount, but at LC it is expected that the effects of beamstralung or bremstralung will not significantly change the available center-of-mass energy of the $e^-e^+$ system [3]. Hence in events generated from the QED process $e_R^- e^+ \to \gamma\gamma$, including possible initial state radiation, the sum of the two photon energies is close to $\sqrt{S}$. By demanding that the sum of the two photon energies is around $\sqrt{S}/2$, *i.e.*, large missing energy in the event, and that the two photons are not back-to-back, one can eliminate the large background from photon pair production.

Another background process at LC is $e_R^- e^+ \to \gamma\gamma Z$ with the $Z$-boson decay $Z \to \nu\bar\nu$ (with three possible neutrino flavors). This process has large missing energy carried away by the neutrino pair. We have calculated the cross-section for this background process. The cross-section diverges at low transverse momentum $p_T$ of either photon, so we require $p_T$ to be larger than $20\,\text{GeV}$ for each photon. We find that this background cross-section is $14.7\,\text{fb}$ (including three neutrino flavors from $Z$-boson decays) which is rather small (*cf.* Table 1). For comparison, with the same $p_T$ cut on the photons the production cross-section of the signal event $e_R^- e^+ \to \chi_1^0 \chi_1^0 \to \gamma\gamma \widetilde{G}\widetilde{G}$ is $210\,\text{fb}$ for $m_{\chi_1^0} = 100\,\text{GeV}$ and $m_{\widetilde{e}_R} = 300\,\text{GeV}$, and it is $16.9\,\text{fb}$ for $m_{\chi_1^0} = 200\,\text{GeV}$ and $m_{\widetilde{e}_R} = 600\,\text{GeV}$. Although the $\gamma\gamma Z$ cross-section is comparable to the signal cross-section in the limit of large SUSY-partner masses, this background event can be easily distinguished from the signal event: The missing energy in the $\gamma\gamma Z$ event is entirely due to the decay of the $Z$-boson, so the invariant mass of the invisible particles (the $\nu\bar{\nu}$ pair) is peaked at the $Z$-boson mass. The invariant mass can be determined from the visible energy, *i.e.*, the initial state and the two photons. The invariant-mass peak for the $\gamma\gamma Z$ background event is shown in Fig. 6, where the invariant mass squared is defined by $M_{inv}^2 = (p_{e^-} + p_{e^+} - p_{\gamma_1} - p_{\gamma_2})^2$. Also shown in Fig. 6 is the distribution in $M_{inv}$ for signal events with $m_{\chi_1^0} = 200\,\text{GeV}$ and $m_{\tilde{e}_R} = 600\,\text{GeV}$, which is a broad distribution. Requiring $M_{inv}$ to be away from the $Z$-peak to discriminate against the $\gamma\gamma Z$ events would further suppress this already small background. For instance, as seen in Fig. 6, if we require $|M_{inv} - m_Z| > 20\,\text{GeV}$, then the ratio of signal to background becomes $14.6\,\text{fb}/0.7\,\text{fb}=21$, as compared to $16.9\,\text{fb}/14.7\,\text{fb}=1.15$ without the $M_{inv}$-cut.

Figure 7 shows the $\chi_1^0\chi_1^0$ production cross-section $\sigma$ for a few choices of $m_{\tilde{e}_R}$, as a function of $m_{\chi_1^0}$ at LC. The order of magnitude of $\sigma$ is 100 fb. We conclude that it is possible to observe the non-background-free signal event at LC with 50 fb$^{-1}$ integrated luminosity, for all the shaded region shown in Fig. 2 except if $m_{\chi_1^0}$ is extremely close to the threshold for $\chi_1^0\chi_1^0$ production, which is $\sqrt{S}/2 = 250\,\text{GeV}$ for LC.

Before closing this section, we comment on the reason for using a right-hand polarized $e^-$ beam for the proposed experiment. The rate of the background process just discussed, $e_R^- e^+ \to \gamma\gamma Z$ with $Z \to \nu\bar{\nu}$, would be almost the same for right-hand polarized electrons or unpolarized electrons: The $e$-$e$-$Z$ coupling is approximately pure axial vector (because the vector part is proportional to $1 - 4\sin^2\theta_w$, which is small) so left-handed or right-handed electrons give the same rate. Similarly, the rates for right-handed electrons or unpolarized electrons would be equal in the case of the QED process $e_R^- e^+ \to \gamma\gamma$, because the electromagnetic coupling is pure vector. In the case of the signal process, the rates for right-handed electrons or unpolarized electrons would be somewhat different: $e_R^- e^+ \to \chi_1^0 \chi_1^0$ occurs by $\widetilde{e_R}$ exchange producing the $\widetilde{\gamma}$ component of $\chi_1^0$ (*cf.* Eq. (3)), whereas $e_L^- e^+ \to \chi_1^0 \chi_1^0$ occurs by $\widetilde{e_L}$ exchange producing both the $\widetilde{\gamma}$ and $\widetilde{W^3}$ components of $\chi_1^0$, and the masses of $\widetilde{e_R}$ and $\widetilde{e_L}$ may be somewhat different. However, we do not expect the signal rates for $e_R^-$ and $e_L^-$ to be very different. So why do we prefer a right-hand polarized $e^-$ beam for our study? One reason for using $e_R^-$ is that the signal process is less model-dependent for $e_R^-$ than $e_L^-$, because the $\widetilde{W^3}$ interactions do not contribute for $e_R^-$. A second reason is that if the $e^-$ is not right-handed, then additional backgrounds must be considered, coming from $e$-$\nu_e$-$W$ interactions. The additional contribution to the background process $e^- e^+ \to \nu_e \bar{\nu}_e \gamma \gamma$ arises from Feynman diagrams in which a t-channel $W$-boson is exchanged between the two fermion lines. (The complete gauge-invariant set of diagrams for this process also includes those diagrams which contribute to $e^- e^+ \to \gamma \gamma Z(\to \nu_e \bar{\nu}_e)$.) This additional contribution vanishes if the $e^-$ beam is right-hand polarized because the $e$-$\nu_e$-$W$ coupling is purely left-handed. Therefore a right-hand polarized electron beam can have a significantly better ratio of signal to backgrounds, compared to an unpolarized $e^-$ beam. Predicting this additional background due to $W$-boson interactions for LC with less than 100% right-hand polarization of the electron beam, or for current colliders which have smaller polarization capability, is a topic for a more detailed study.

## 2.5 Detecting a Light Gravitino From the Decay of a Higgsino-Like NLSP

So far we have only discussed the case that $\chi_1^0$ is gaugino-like. In that case, with a right-hand polarized $e^-$ beam, the production of $\chi_1^0 \chi_1^0$ is related to the production of a photino pair, by Eq. (3). Also, in that case the branching ratio for the decay of the photino component of $\chi_1^0$ into photon plus gravitino is nearly 100% at tree level. Thus our predictions if $\chi_1^0$ is gaugino-like depend on only a few parameters, *i.e.*, $\alpha$, $M_{susy}$, $m_{\chi_1^0}$ and $m_{\widetilde{e}_R}$. Now we consider the case that $\chi_1^0$ is Higgsino-like. As pointed out in Sec. 2.1, in this case only the s-channel diagram (in which $e^-e^+$ annihilate through a virtual Z-boson into $\chi_1^0\chi_1^0$) contributes to $\chi_1^0\chi_1^0$ production, and both the production rate and the decay branching ratios of $\chi_1^0$ depend on the detailed values of SUSY parameters such as $\mu$, $\tan\beta$, $M_1$, and $M_2$ [9], in addition to the previous parameters. Since $\chi_1^0$ is the NLSP and the gravitino is the LSP in the class of models considered [1], at tree level $\chi_1^0$ will decay into $h^0\widetilde{G}$ if the lightest Higgs boson $h^0$ is lighter than $\chi_1^0$. (Here, for simplicity we assume that $R$-parity is conserved and that the other scalar fields in the SUSY models are heavier than $\chi_1^0$.) If this decay channel is allowed, then the branching ratio for $\chi_1^0 \to \gamma\widetilde{G}$ will be small because the latter decay can only occur via loop corrections. Even if $m_{\chi_1^0} < m_{h^0}$, so that the tree-level decay channel is forbidden, the branching ratio for $\chi_1^0 \to \gamma\widetilde{G}$ is still not 100% if the one-loop process $\chi_1^0 \to Z\widetilde{G}$ is possible, *i.e.*, if $m_{\chi_1^0} > m_Z$. The branching ratio for a specific decay mode will depend on the masses of top-quark, top-squark (stop), bottom-squark (sbottom), $\tan\beta$, *etc.*

In this paper we shall not carry out a detailed study for the case that $\chi_1^0$ is Higgsino-like, because any conclusion will depend strongly on the details of the SUSY parameters. Nevertheless, we speculate that the detection of the gravitino in the Higgsino-like case via the decay mode $\chi_1^0 \to \gamma\widetilde{G}$ will be relatively more difficult than that in the gaugino-like case. Because a one-loop process typically has a suppression factor $1/(16\pi^2) \approx 10^{-2}$ in the decay rate compared to a tree-level process, the lifetime for $\chi_1^0 \to \gamma\widetilde{G}$ in the Higgsino-like case will typically be longer by a factor of order $10^2$ than that in the gaugino-like case. A Higgsino-like $\chi_1^0$ will travel a longer distance, by a factor of order $10^2$, than a gaugino-like $\chi_1^0$, before decaying to $\gamma\widetilde{G}$. Hence, as discussed in previous sections (*cf.* Table 1) it is less likely that this decay can be observed, because the decay will more likely occur outside the detector volume. From these considerations we expect that for a large range of $M_{susy}$ and $m_{\chi_1^0}$, detecting the gravitino via the process $e_R^- e^+ \to \chi_1^0(\to \gamma \widetilde{G})\chi_1^0(\to \gamma \widetilde{G})$ will not be guaranteed at LC if $\chi_1^0$ is Higgsino-like. A further, more detailed study would be necessary to determine accurately the accessible range of the parameters $M_{susy}$ and $m_{\chi_1^0}$ for this case.

In the case that $\chi_1^0$ is Higgsino-like with $m_{\chi_1^0} > m_{h^0}$, it is more natural to study the tree-level decay mode $\chi_1^0 \to h^0 \widetilde{G}$, where $h^0$ is the lightest Higgs boson. Assuming $m_{h^0} < 2m_t \simeq 350\,\mathrm{GeV}$, the lightest Higgs boson $h^0$ will most likely decay to a $b\bar{b}$ pair. Hence, the signature of the signal would be 4 $b$-jets plus missing energy. To suppress the large backgrounds of the form $e^- e^+ \to 4$ jets plus missing energy, a good efficiency in $b$-tagging is needed. If the decay of $\chi_1^0 \to h^0 \widetilde{G}$ occurs inside the detector volume and the two $h^0$-bosons can be reconstructed, then $e^- e^+ \to \chi_1^0(\to h^0 \widetilde{G})\chi_1^0(\to h^0 \widetilde{G})$ will be the dominant process to produce a pair of $h^0$-bosons for models in which $\widetilde{G}$ is the LSP and $\chi_1^0$ is the NLSP. Thus, this event signature is also unique for models with low $M_{susy}$ and deserves a detailed study of its own, which is beyond the scope of this paper.

## 3   Discussion and Conclusion

We have discussed the possibility of detecting a light gravitino at LC, a proposed future $e^- e^+$ collider with center-of-mass energy $\sqrt{S} = 500\,\mathrm{GeV}$ and with a right-hand polarized $e^-$ beam, from the decay $\chi_1^0 \to \gamma \widetilde{G}$ where the neutralino $\chi_1^0$ is produced in pairs in the $e^- e^+$ collisions. The proposed LC would also operate at $\sqrt{S} = 1$ or $1.5\,\mathrm{TeV}$ with a luminosity of $200\,\mathrm{fb}^{-1}$ per year. Figure 8 shows the total cross-section $\sigma(e_R^- e^+ \to \chi_1^0 \chi_1^0)$ for a gaugino-like $\chi_1^0$ with $\alpha = 0$, as a function of $\sqrt{S}$, for two cases of the mass parameters: $m_{\chi_1^0} = 100\,\mathrm{GeV}$ $m_{\widetilde{e}_R} = 300\,\mathrm{GeV}$, and $m_{\chi_1^0} = 200\,\mathrm{GeV}$ $m_{\widetilde{e}_R} = 600\,\mathrm{GeV}$. If $m_{\chi_1^0}$ is large then the cross-section increases with $\sqrt{S}$; but if $m_{\chi_1^0} = 100\,\mathrm{GeV}$ then the cross-section actually decreases as $\sqrt{S}$ increases from $500\,\mathrm{GeV}$ to $1\,\mathrm{TeV}$. The ranges of the parameters $M_{susy}$ and $m_{\chi_1^0}$ that are accessible for the three modes of LC are shown in Fig. 9. (For simplicity, we did not separate the regions of parameters for background-free signal events from those for non-background-free signal events, as we did in Fig. 2.) It is clearly seen that a 500 GeV LC will probe a slightly larger region of the parameters $M_{susy}$ and $m_{\chi_1^0}$ than a TeV LC if $m_{\chi_1^0} < 250$ GeV. (At higher $\sqrt{S}$ the neutralinos are more likely to exit the detector before decaying.) Therefore, a TeV LC is needed to detect a light gravitino through the event signature of 2 photons plus missing energy, only if the NLSP $\chi_1^0$ is heavier than about 250 GeV.

We should also consider whether this process can be discovered at a current $e^-e^+$ or $p\bar{p}$ collider. First consider the case of LEP/SLC, with center-of-mass energy $\sqrt{S} = m_Z = 91$ GeV and integrated luminosity $450\,\mathrm{pb}^{-1}$ (which is about the integrated luminosity at $m_Z$ when combining all the experiments from LEP and SLC), and also LEP-II, with center-of-mass energy $\sqrt{S} = 190$ GeV and luminosity $500\,\mathrm{pb}^{-1}$ per year per experiment. Figure 10 shows the range of parameters $M_{susy}$ and $m_{\chi_1^0}$ accessible by LEP/SLC, LEP-II, and LC, superimposed, with the same assumptions as Fig. 2. The accessible region depends mainly on kinematics, *i.e.*, on the requirement that $D < 1$ m, where $D$ is the distance traveled by the neutralino before it decays, and on the threshold mass $m_{\chi_1^0} = \sqrt{S}/2$. In the case of a gaugino-like $\chi_1^0$ at LC we found that the cross-section for right-handed electrons is of order 100 fb (*cf.* Table 1), large enough to discover the signal even for large $m_{\widetilde{e}_R}$ and $m_{\chi_1^0}$ near the threshold. In the case of LEP/SLC or LEP-II with unpolarized $e^-$ beam, the cross-section depends on the masses of both selectrons ($\widetilde{e}_R$ and $\widetilde{e}_L$), but if we assume these masses are about equal then we may *estimate* that the unpolarized cross-section is approximately equal to the right-handed $e^-$ cross-section. The boundaries for LEP/SLC and LEP-II in Fig. 10 were calculated with this assumption, with $m_{\widetilde{e}_R} = m_{\widetilde{e}_L} = 300$ GeV. For example, the cross-section for LEP-II ($\sqrt{S} = 190\,\text{GeV}$) with $m_{\chi_1^0} = 100\,\text{GeV}$ and $m_{\widetilde{e}_R} = m_{\widetilde{e}_L} = 300\,\text{GeV}$ is 143 fb. In Fig. 10 the triangular area enclosed by the curve for each collider is the range of parameters such that there would be 10 events (total for LEP/SLC, or per year per experiment for LEP-II or LC), with decay length $D < 1\,\text{m}$.

Figure 10 implies that for $M_{susy}$ between $10^5\,\text{GeV}$ and $10^7\,\text{GeV}$, the range relevant to the model of Ref. [1], only a very limited region of parameter space is accessible at LEP/SLC, but an interesting region for $M_{susy} < 10^6\,\text{GeV}$ will be accessible at LEP-II. However, because the electron beam polarization at LEP-II is only of order 50%, there is a background from left-handed electrons: the process $e^-e^+ \to \nu_e \bar{\nu}_e \gamma\gamma$ in which a $W$-boson is exchanged between the fermions. This background may be significant at the level of 10 events. A definitive analysis of a LEP-II search for the gravitino process must include this background.

Next we turn to the Tevatron $p\bar{p}$ collider. We have *estimated* the cross-section for the process $p\bar{p} \to \chi_1^0 \chi_1^0$ at $\sqrt{s} = 2$ TeV, for gaugino-like $\chi_1^0$, using a Monte Carlo program with CTEQ2 parton distribution functions. We estimate that the cross-section is 38 fb for $m_{\chi_1^0} = 100\,\text{GeV}$ and $m_{\widetilde{q}} = 100\,\text{GeV}$, where $\widetilde{q}$ indicates any squark. The cross-section decreases with $m_{\widetilde{q}}$. Because of the small production rate (about a factor of 15 smaller than a 500 GeV LC) and the additional large backgrounds in hadron collisions (either from physics processes or from the imperfectness of the detector), the current total integrated luminosity of the Tevatron is probably too small to provide a useful search for the light gravitino. With the upgrade of the Tevatron, at which the luminosity will increase by an order of magnitude (to about $2\,\text{fb}^{-1}$ per year), one can probably detect the light gravitino for some values of $M_{susy}$ and $m_{\chi_1^0}$. This requires a separate study as well.

In conclusion, the proposed LC can in the future provide a means to search for the gravitino decay of the NLSP $\chi_1^0$, at least if $\chi_1^0$ is gaugino-like, for a significant part of the parameter space relevant to models in which the SUSY-breaking scale is low, *i.e.*, $M_{susy}$ within a few orders of magnitude of the weak scale. The $\chi_1^0\chi_1^0$ production cross-section is large enough, of order 100 fb, that it is not a limiting factor. The limiting factor is the lifetime of the NLSP. The lifetime is proportional to $M_{susy}^4$. If $M_{susy}$ is too large, then the NLSP will exit the detector before decaying, and no information on the gravitino will be obtained. But as shown in Fig. 2, if $M_{susy}$ is small enough for a given neutralino mass $m_{\chi_1^0}$, then the neutralinos will decay inside the LC detector. The decay $\chi_1^0 \to \gamma \widetilde{G}$ can then be used to detect the gravitino by seeing the two photons with large missing energy from $\chi_1^0$ pair production.

## 4 Acknowledgements

We would like to thank Michael Dine, Jonathan Feng, Ray Frey, Joey Huston, Gordon L. Kane, Jim Linnemann, Eric Poppitz, Wayne Repko, Byron Roe, and Carl Schmidt for helpful discussions. The work of C.–P.Y. was supported in part by NSF grant #PHY-9309902.

# Tables

Table 1: Production cross-section $\sigma$ and typical decay distance $D$ for the process $e_R^- e^+ \to \chi_1^0 \chi_1^0 \to \gamma \widetilde{G} \gamma \widetilde{G}$ at LC, assuming $\alpha = 0$, for several values of masses. These numbers are for $M_{susy} = 10^6$ GeV.

| $m_{\chi_1^0}$ (GeV) | $m_{\widetilde{e}_R}$ (GeV) | $\sigma$ (fb) | $D$ (m) |
|---|---|---|---|
| 100 | 100 | 670 | 0.35 |
| 100 | 300 | 252 | 0.35 |
| 100 | 600 | 57.1 | 0.35 |
| 100 | 1000 | 11.7 | 0.35 |
| 200 | 300 | 84.6 | 0.0036 |
| 200 | 600 | 17.6 | 0.0036 |
| 200 | 1000 | 3.4 | 0.0036 |

# Figures

Figure 1: Differential cross-section $d\sigma/d\cos\theta$ for $e_R^- e^+ \to \chi_1^0 \chi_1^0$, where $\theta$ is the angle of a neutralino, at $\sqrt{S} = 500\,\text{GeV}$. The mixing angle $\alpha$ is 0, and the masses are $m_{\chi_1^0} = 100\,\text{GeV}$, $m_{\widetilde{e}_R} = 300\,\text{GeV}$ (solid curve), and $m_{\chi_1^0} = 200\,\text{GeV}$, $m_{\widetilde{e}_R} = 600\,\text{GeV}$ (dashed curve).

Figure 2: Range of parameters $M_{susy}$ and $m_{\chi_1^0}$ accessible at LC. The shaded region is the range of $M_{susy}$ and $m_{\chi_1^0}$ for which $D < 1\,\text{m}$ and $\sigma > 0.2$ fb, where $D$ is the typical decay length of $\chi_1^0$ and $\sigma$ is the production cross-section for $e_R^- e^+ \to \chi_1^0 \chi_1^0$ at LC. The bound $\sigma > 0.2$ fb is equivalent to observing more than 10 events assuming integrated luminosity 50 fb$^{-1}$. The cross-shaded region is for $10\,\text{cm} < D < 1\,\text{m}$, corresponding to the background-free signal process. Parameter values for this plot are $\alpha = 0$ and $m_{\widetilde{e}_R} = 300\,\text{GeV}$.

Figure 3: Differential cross-section in photon energy $d\sigma/dE_\gamma$ for $e_R^- e^+ \to \chi_1^0 \chi_1^0 \to \gamma \widetilde{G} \gamma \widetilde{G}$ at LC, assuming $\alpha = 0$. The mass parameter values are $m_{\chi_1^0} = 100$ GeV and $m_{\widetilde{e}_R} = 300$ GeV.

Figure 4: Differential cross-section in photon transverse momentum $d\sigma/dp_T^\gamma$ for $e_R^- e^+ \to \chi_1^0 \chi_1^0 \to \gamma \tilde{G} \gamma \tilde{G}$ at LC, assuming $\alpha = 0$. The mass parameter values are $m_{\chi_1^0} = 100$ GeV and $m_{\tilde{e}_R} = 300$ GeV.

Figure 5: Differential cross-section in the sum of photon energies $d\sigma/d(E_{\gamma_1} + E_{\gamma_2})$ for $e_R^- e^+ \to \chi_1^0 \chi_1^0 \to \gamma \tilde{G} \gamma \tilde{G}$ at LC, assuming $\alpha = 0$. The mass parameter values are $m_{\chi_1^0} = 100$ GeV and $m_{\tilde{e}_R} = 300$ GeV.

Figure 6: Differential cross-section $d\sigma/dM_{inv}$ in the invariant mass of invisible particles $M_{inv}$ for the signal process $e_R^- e^+ \to \chi_1^0 \chi_1^0 \to \gamma \tilde{G} \gamma \tilde{G}$, and for the background process $e_R^- e^+ \to \gamma\gamma Z$, with decay $Z \to \nu\bar{\nu}$. The photon transverse momenta are required to be greater than 20 GeV. The masses for the signal process are $m_{\chi_1^0} = 200$ GeV and $m_{\tilde{e}_R} = 600$ GeV.

Figure 7: Total cross-section at LC for the process $e_R^- e^+ \to \chi_1^0 \chi_1^0$ with $\alpha = 0$, as a function of $m_{\chi_1^0}$ for $m_{\tilde{e}_R} = 300$ GeV (solid curve), $m_{\tilde{e}_R} = 100$ GeV (dashed curve), and $m_{\tilde{e}_R} = 600$ GeV (dot dashed curve). (Note that only $m_{\chi_1^0} < m_{\tilde{e}_R}$ is allowed in the models we consider because $\chi_1^0$ is the NLSP.)

Figure 8: Total cross-section for the process $e_R^- e^+ \to \chi_1^0 \chi_1^0$ with $\alpha = 0$, as a function of $\sqrt{S}$. The mass parameter values are $m_{\chi_1^0} = 100$ GeV, $m_{\tilde{e}_R} = 300$ GeV (solid curve), and $m_{\chi_1^0} = 200$ GeV, $m_{\tilde{e}_R} = 600$ GeV (dashed curve).

Figure 9: Range of parameters $M_{susy}$ and $m_{\chi_1^0}$ accessible at LC with $\sqrt{S} = 500$ GeV (solid line), 1 TeV (dashed line), and 1.5 TeV (dot-dash line). The luminosity per year is 50, 200 and 200 fb$^{-1}$, respectively. Mass parameter values are $m_{\tilde{e}_R} = 300$ GeV, 500 GeV and 750 GeV, respectively, and the mixing angle $\alpha$ is 0. The interior of each triangle is the region where the number of events is greater than 10, and $D < 1$ m.

Figure 10: Range of parameters $M_{susy}$ and $m_{\chi_1^0}$ accessible at LC (solid line), LEP/SLC (dashed line), and LEP-II (dot-dash line). The interior of each triangle is the region where the number of events is greater than 10, and $D < 1$ m. Parameter values for this plot are $\alpha = 0$ and $m_{\tilde{e}_R} = 300$ GeV.

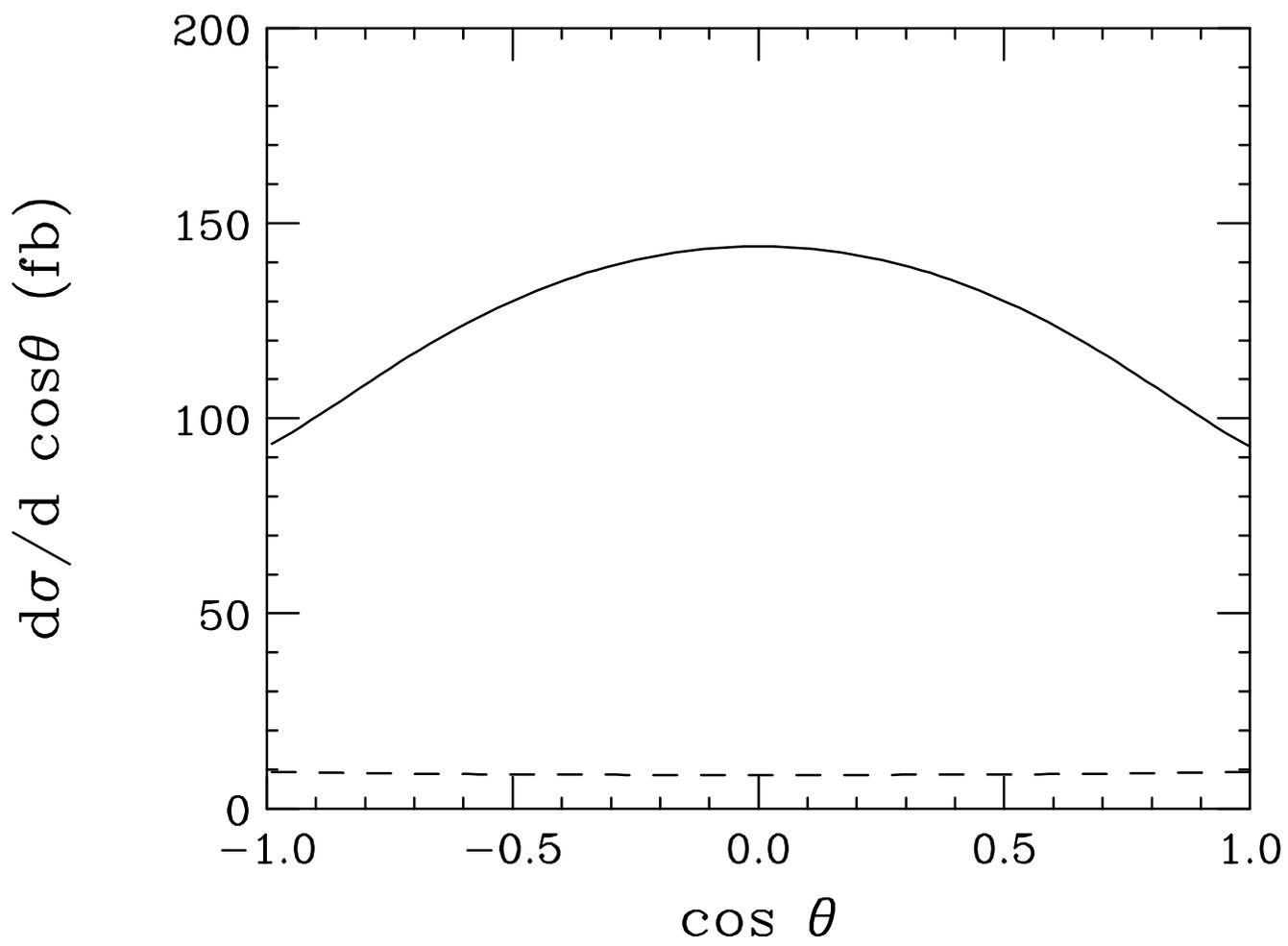

*Figure 1*

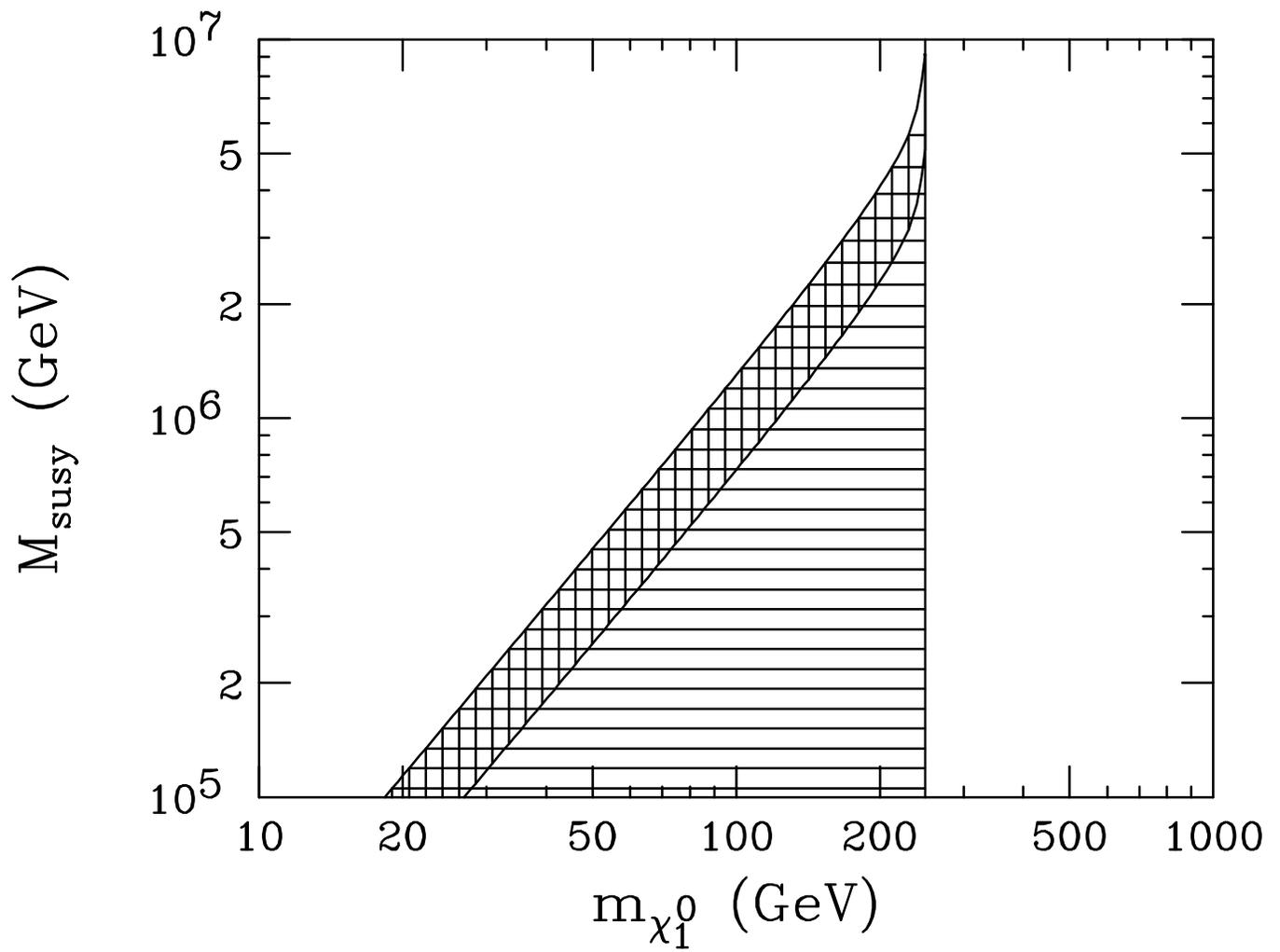

*Figure 2*

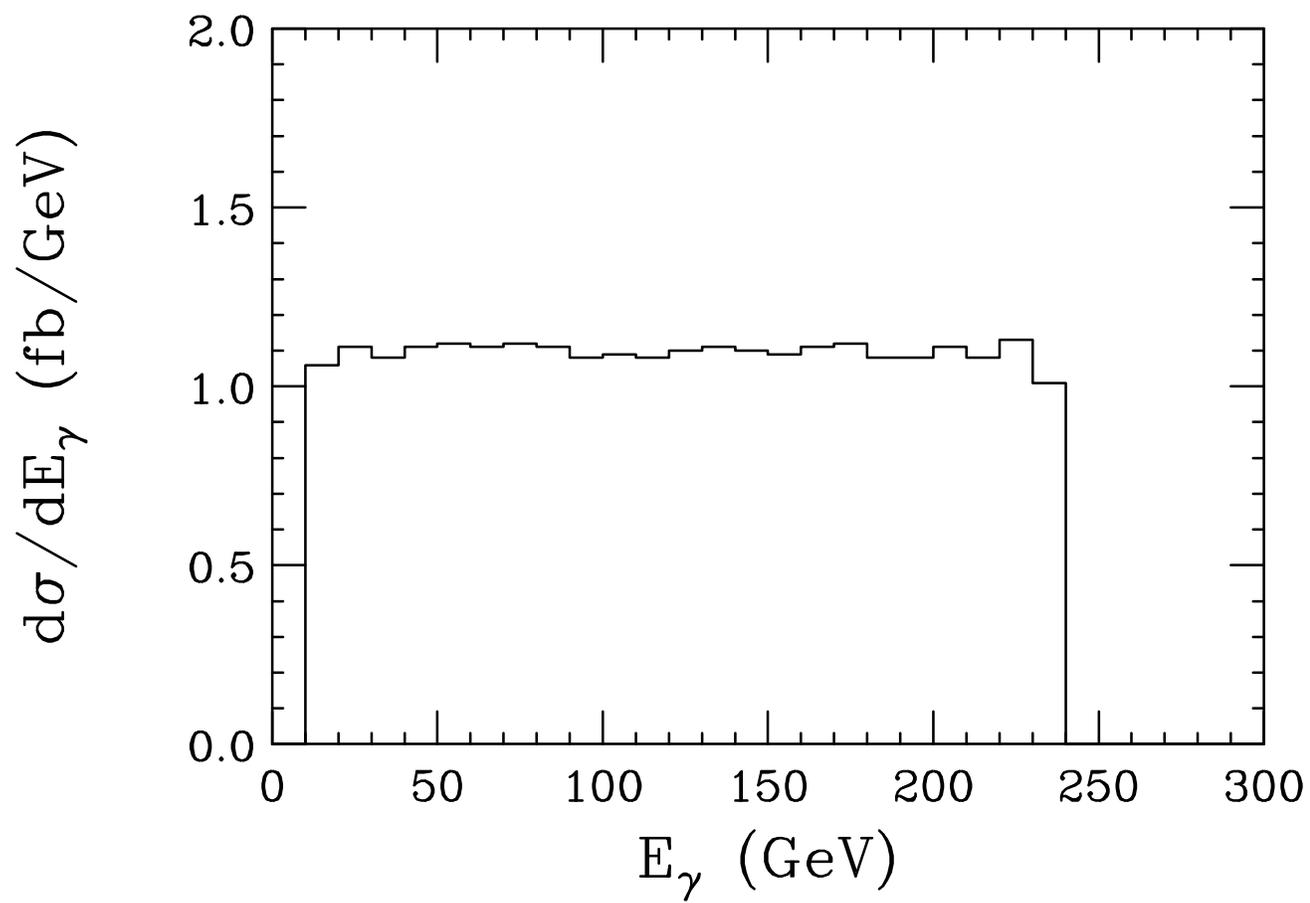

*Figure 3*

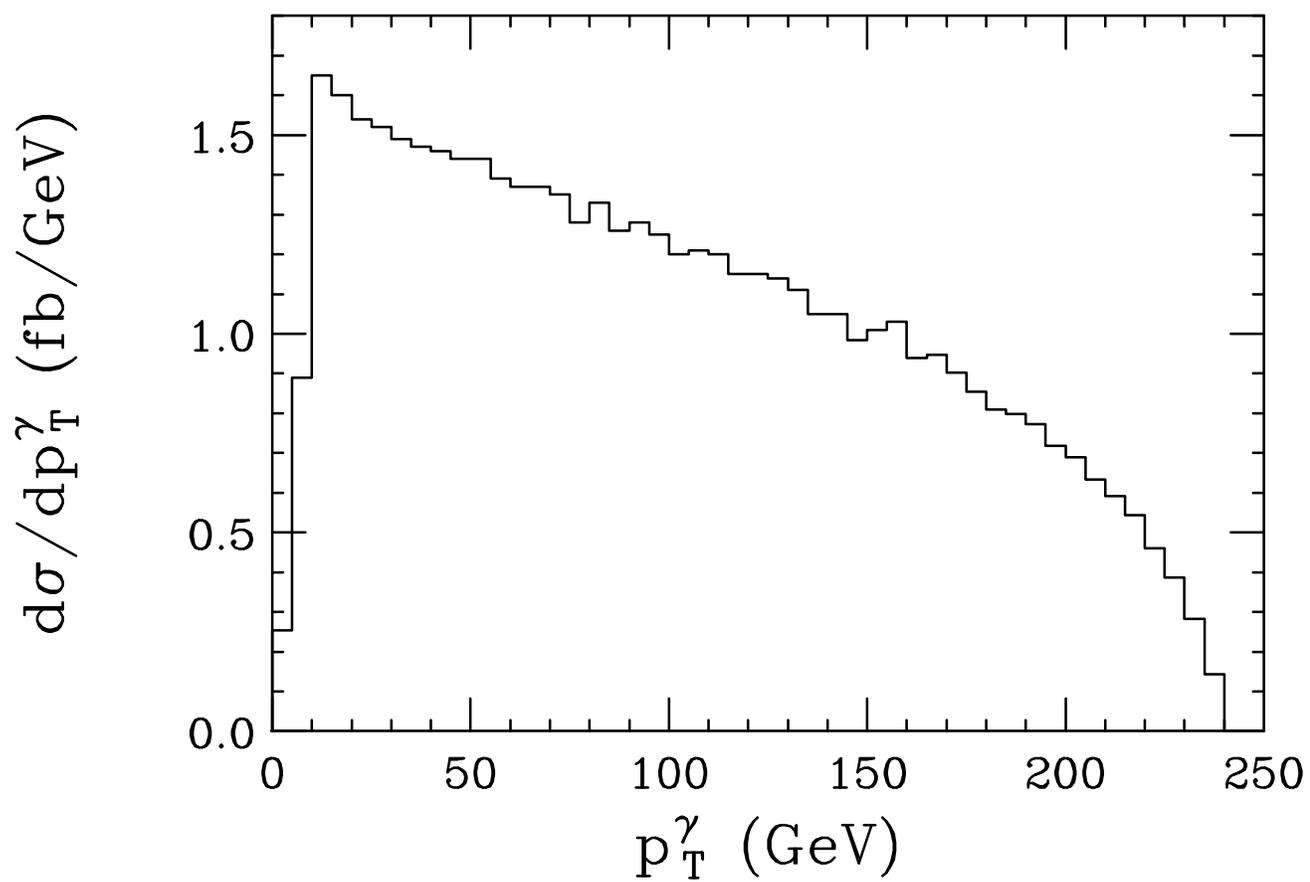

*Figure 4*

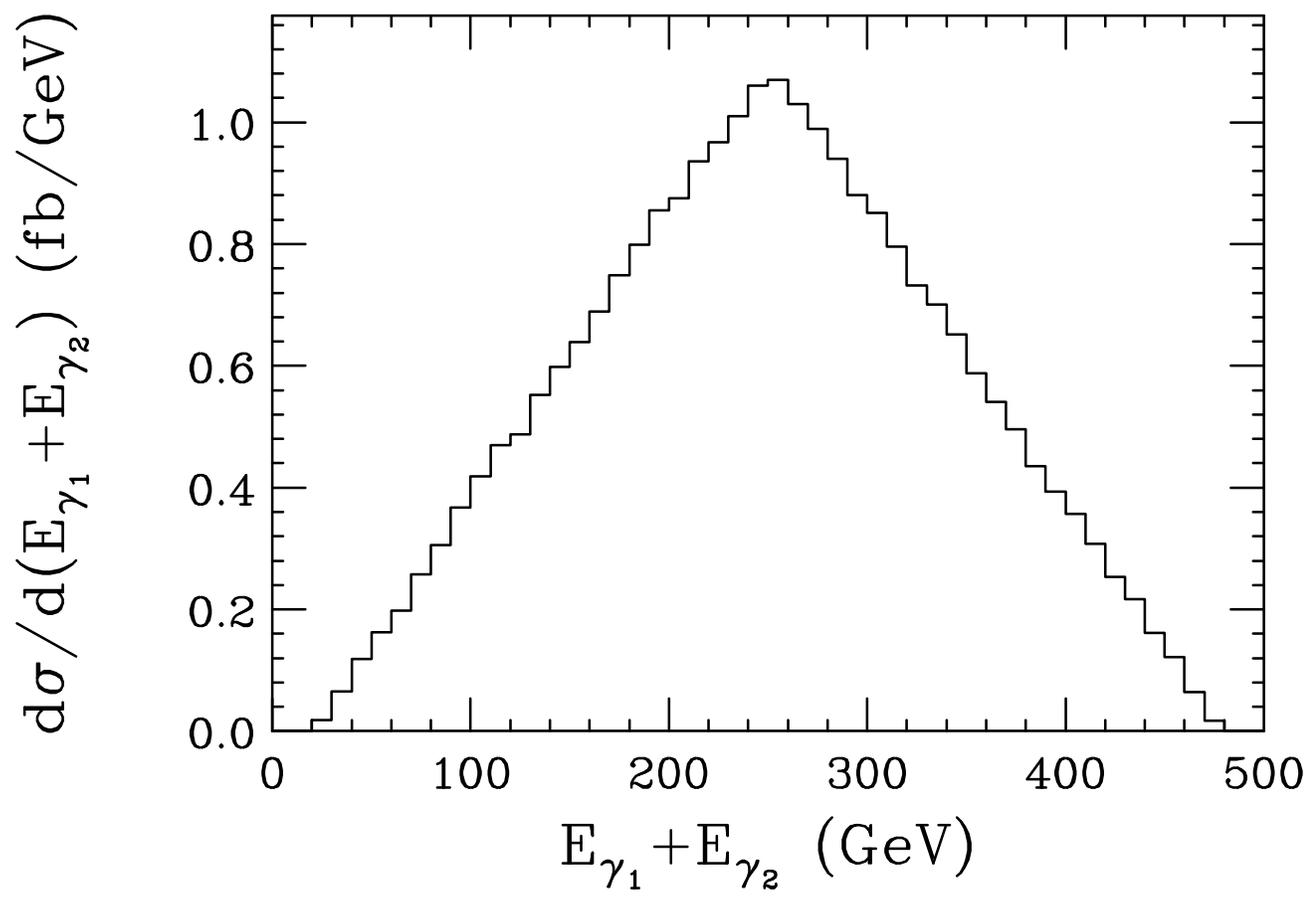

*Figure 5*

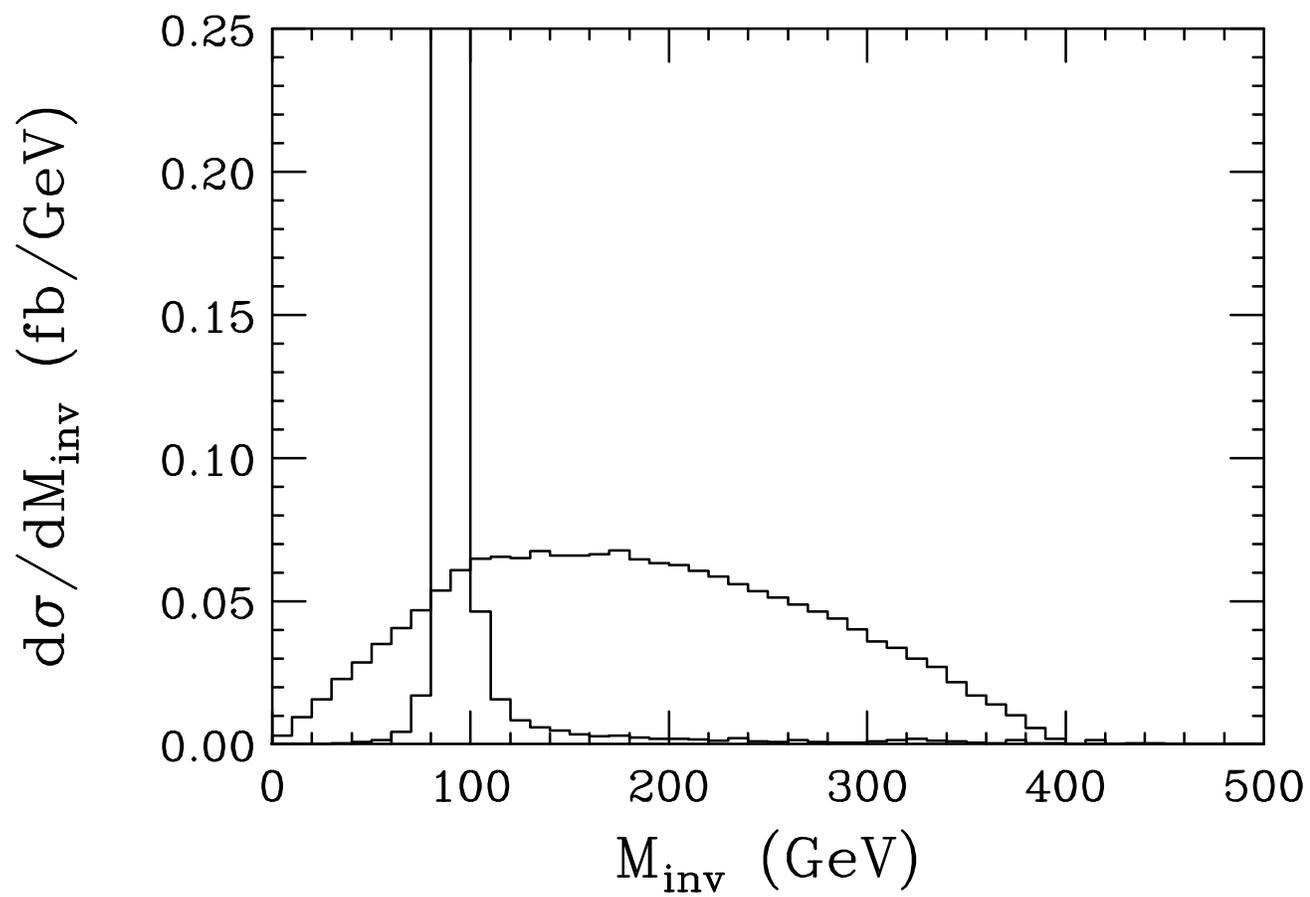

*Figure 6*

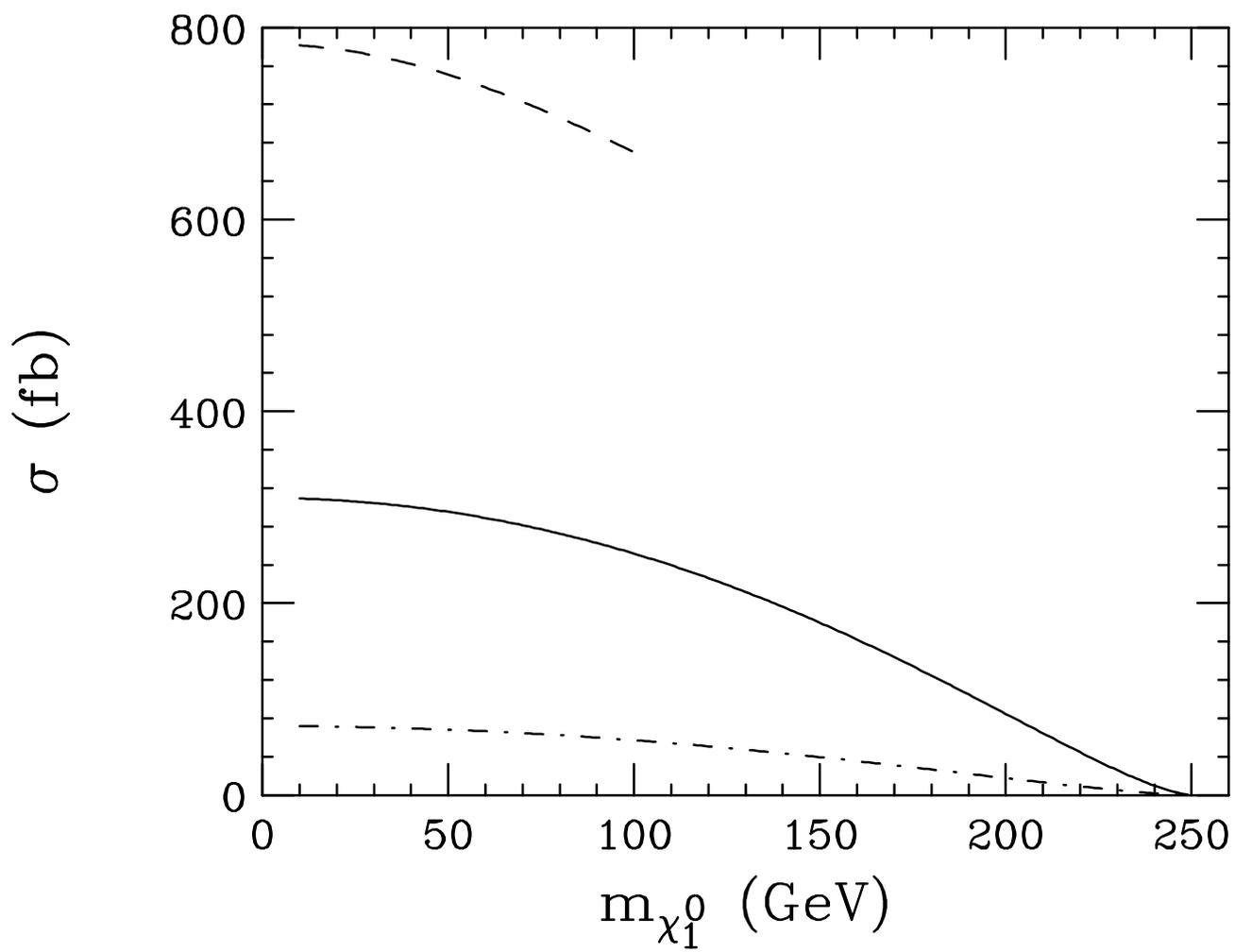

*Figure 7*

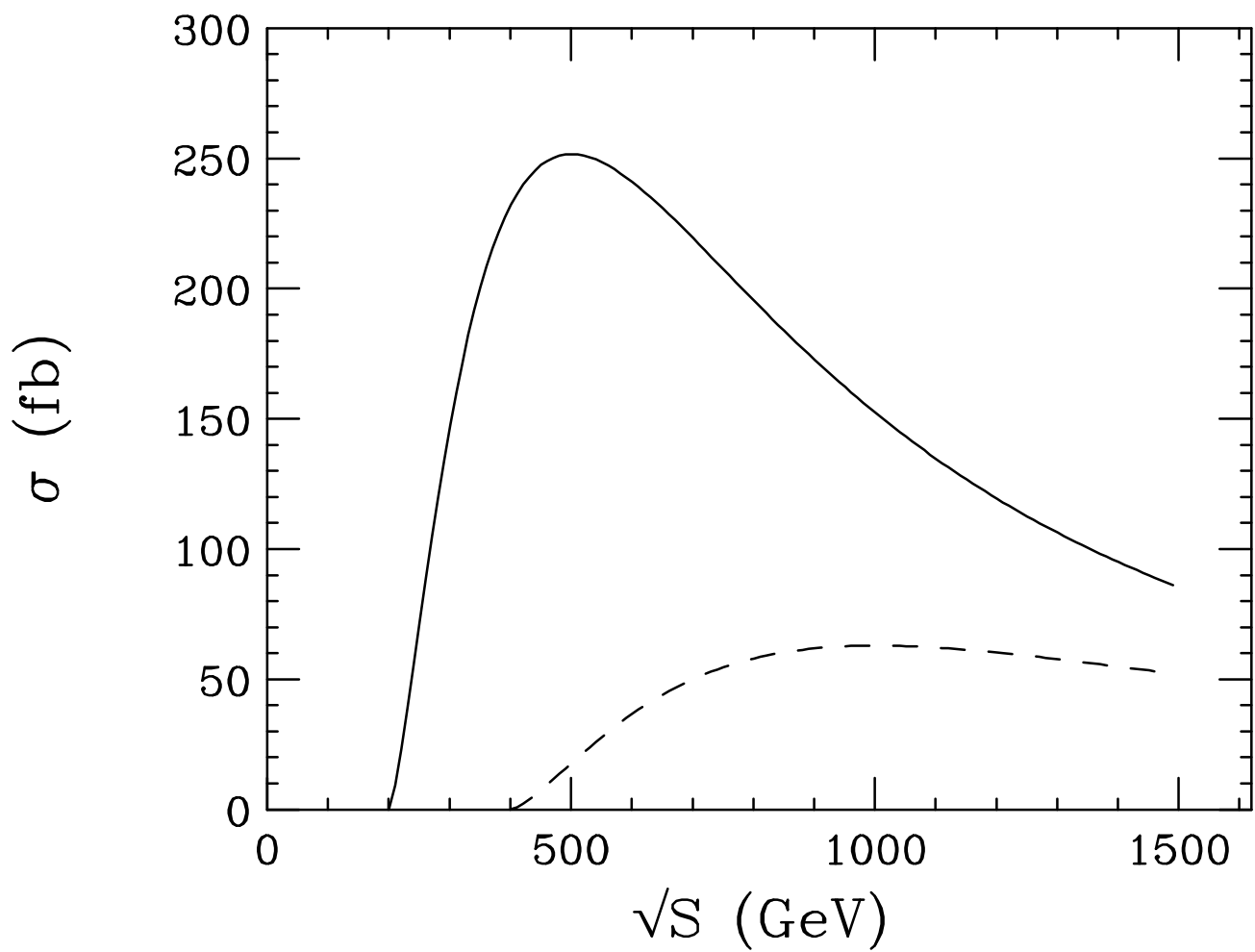

*Figure 8*

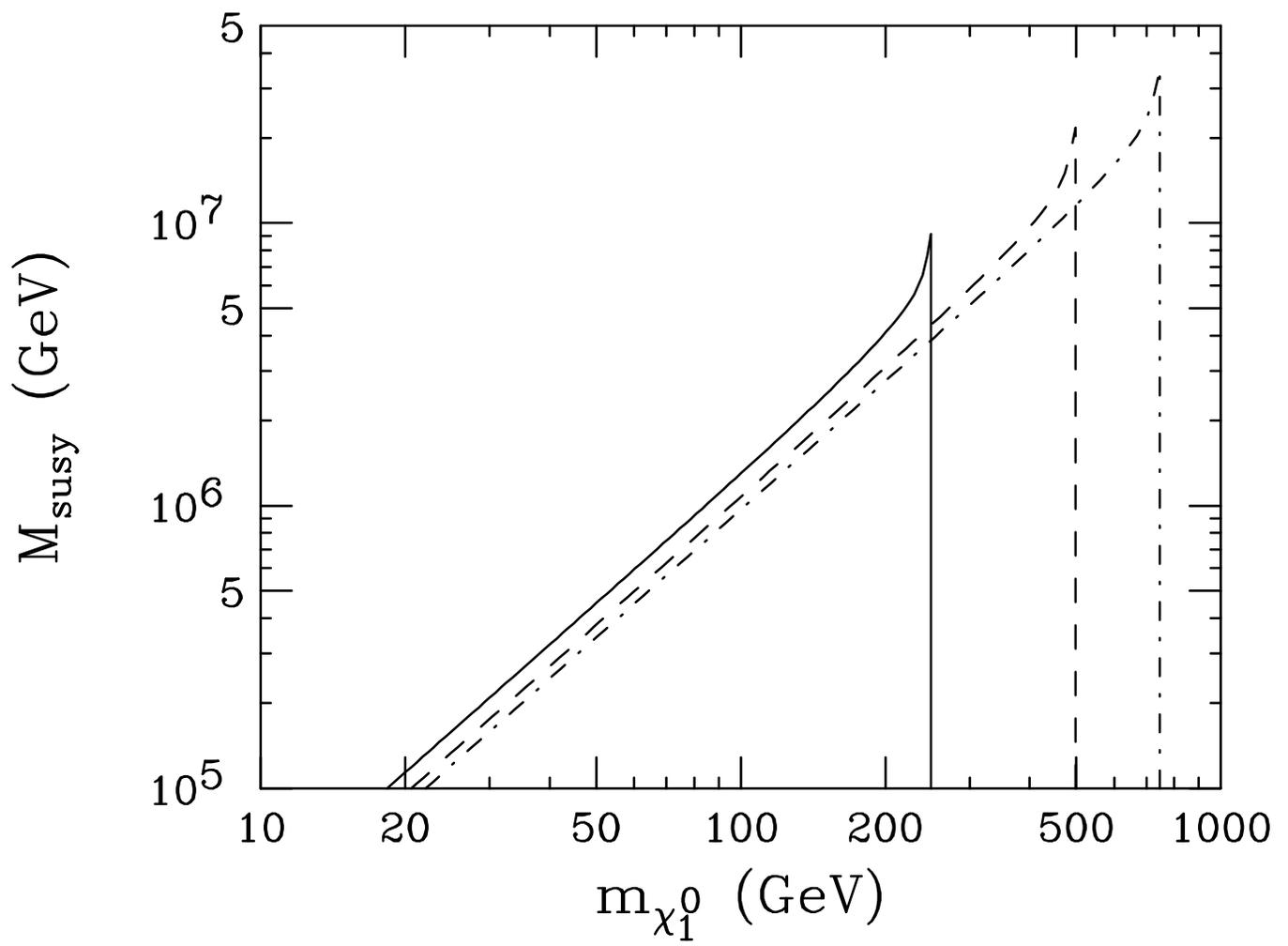

*Figure 9*

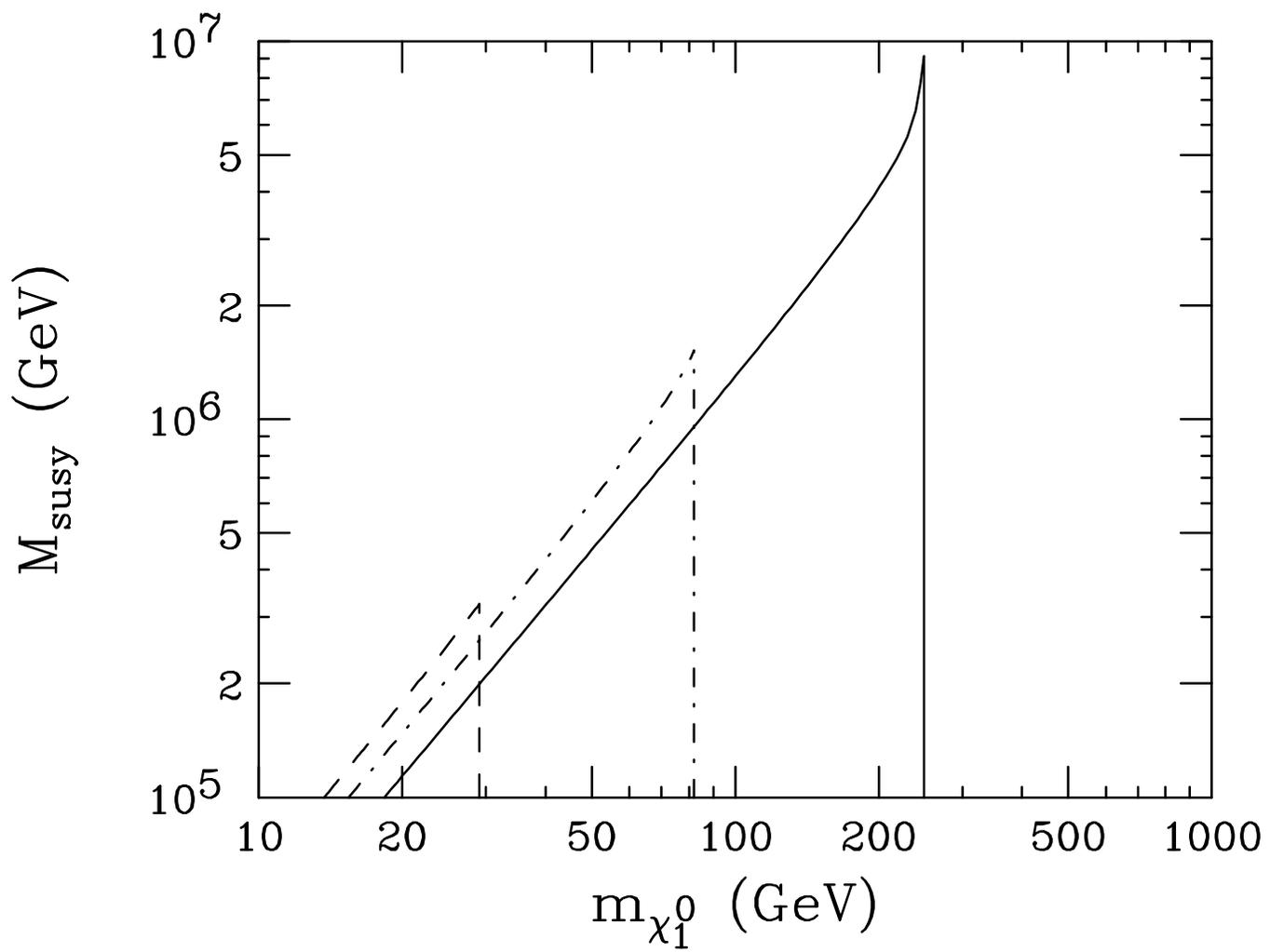

*Figure 10*